\documentclass{article}

\usepackage{PRIMEarxiv}

\usepackage{hyperref}       
\usepackage{url}            
\usepackage{booktabs}       
\usepackage{amsfonts}       
\usepackage{nicefrac}       
\usepackage{microtype}      
\usepackage{lipsum}
\usepackage{fancyhdr}       
\usepackage{graphicx}       
\graphicspath{{media/}}     

\usepackage{hyperref}
\usepackage[utf8x]{inputenc}
\usepackage[english]{babel}
\usepackage{booktabs}

\usepackage[nodisplayskipstretch]{setspace}
\usepackage{subcaption}

\usepackage{amsmath, amsthm, amssymb}
\usepackage{bbm}
\usepackage[makeroom]{cancel}

\usepackage{gensymb}
\usepackage{multirow}
\usepackage{graphicx}
\usepackage{adjustbox}

\usepackage{float}
\usepackage{algorithm}
\usepackage{algpseudocode}

\usepackage[title]{appendix}

\usepackage{varwidth}
\usepackage{calc}

\usepackage{array,ragged2e}
\usepackage[flushleft]{threeparttable}

\usepackage{relsize}



\makeatletter
\let\oldnl\nl
\newcommand{\nonl}{\renewcommand{\nl}{\let\nl\oldnl}}
\makeatother

\newcounter{algocont}

\makeatletter

\newtheorem{theorem}{Theorem}

\algnewcommand\algorithmicforeach{\textbf{for each}}
\algdef{S}[FOR]{ForEach}[1]{\algorithmicforeach\ #1\ \algorithmicdo}

\title{Exploiting Hierarchical Dependence Structures for Unsupervised Rank Fusion in Information Retrieval
}

\author{
  J. Hermosillo-Valadez\thanks{\textit{Corresponding author}}\\
  Centro de Investigación en Ciencias \\
  Universidad Autónoma del Estado de Morelos \\
  Cuernavaca, México\\
  \texttt{jhermosillo@uaem.mx}\\
  \And
  E. Morales-González\\
  Centro de Investigación en Ciencias \\
  Universidad Autónoma del Estado de Morelos \\
  Cuernavaca, México\\
 \texttt{eliseo.moralesgon@uaem.edu.mx}\\
  \And
  F. Fern\'{a}ndez-Reyes \\
  Centro de Investigación en Ciencias \\
  Universidad Autónoma del Estado de Morelos \\
  Cuernavaca, México\\
 \texttt{francisfdez83@gmail.com} \\
   \And
  M. Montes-y-G\'{o}mez \\
  Laboratorio de Tecnologías del Lenguaje\\
  Instituto Nacional de Astrofísica, Óptica y Electrónica \\
  Sta María Tonanzintla, San Andrés Cholula, México\\
  \texttt{mmontesg@inaoep.mx} \\
   \And
  J. Fuentes-Pacheco \\
  Centro de Investigación en Ciencias \\
  Universidad Autónoma del Estado de Morelos \\
  Cuernavaca, México\\
  \texttt{jorge.fuentes@uaem.mx} \\
   \And
  J.M. Rendón-Mancha \\
  Centro de Investigación en Ciencias \\
  Universidad Autónoma del Estado de Morelos \\
  Cuernavaca, México\\
  \texttt{rendon@uaem.mx} \\
}

\begin{document}
\maketitle

\begin{abstract}
The goal of rank fusion in information retrieval (IR) is to deliver a single output list from multiple search results. Improving performance by combining the outputs of various IR systems is a challenging task. A central point is the fact that many non-obvious factors are involved in the estimation of relevance, inducing nonlinear interrelations between the data. The ability to model complex dependency relationships between random variables has become increasingly popular in the realm of information retrieval, and the need to further explore these dependencies for data fusion has been recently acknowledged. Copulas provide a framework to separate the dependence structure from the margins. Inspired by the theory of copulas, we propose a new unsupervised, dynamic, nonlinear, rank fusion method, based on a nested composition of non-algebraic function pairs. The dependence  structure of the model is tailored by leveraging query-document correlations on a per-query basis. We experimented with three topic sets over CLEF corpora fusing 3 and 6 retrieval systems, comparing our method against the CombMNZ technique and other nonlinear unsupervised strategies. The experiments show that our fusion approach improves performance under explicit conditions, providing insight about the circumstances under which linear fusion techniques have comparable performance to nonlinear methods.
\end{abstract}


\section{Introduction}
Post-search data fusion consists in obtaining a single result set out of the outputs of different search engines, constituting a suitable mechanism for improving the performance of information retrieval (IR) systems. 
A long history of theoretical and empirical research work has shown that this is a challenging task \cite{Cummins2011,Arampatzis2011,Robertson2013}. 
A central point is the fact that many non-obvious factors are involved in the estimation of relevance, inducing nonlinear interrelations between the data \cite{eickhoff2013}. 

The ability to model complex dependency relationships between random variables has become increasingly popular in the realm of information retrieval, as evidenced by recent new approaches recurring to Machine Learning and Deep Learning strategies \cite{Mitra_INR_2018}, \cite{dai2020DeepCT}, \cite{mallia2021DeepImpact}, and the need to further explore these dependencies for data fusion has been recently acknowledged \cite{canalle_survey_2021}. 

In this paper, we propose a data fusion framework based upon non-algebraic function compositions. The method draws from the theory of copulas, for which supervised methods have been proposed as a way for coping with dependency issues in IR and Data Mining \cite{eickhoff2013, eickhoff2014, komatsuda2016, gorecki_tau_2016}. 

The contributions of our work are the following:
\begin{enumerate}
    \item We show how to combine IR engines' output sets, by combining pairs of nonlinear functions whose parameters are fitted using a concordance measure between retrieval systems. 
    \item The model parameters are further tailored as a function of an estimation of each document's relevance on a per-request basis. As a result, we propose an unsupervised fusion framework, which allows to derive dynamic nonlinear function combinations with no prior knowledge on relevance nor training phases. 
    \item We show that our novel rank fusion approach may improve the overall IR process performance, showing the advantages of nonlinear methods, but akin to linear models which are insensitive to pre-trained parameters. 
    \item Although questions on why and how linear data fusion schemes perform well have been thoroughly discussed in the past \cite{wu2006, frankhsu2005, wu2015geometric}, there is a lack of insight about the limits of performance improvement of nonlinear data fusion techniques over linear ones. Our approach gives insight about the circumstances under which linear data fusion techniques may have a comparable performance to nonlinear methods. 
\end{enumerate}

The remainder of the paper is organized as follows: 
In Section \ref{sec:rel} we review the related work. In Section \ref{sec:pb} we set out the problem and background of our work. We describe our method and fusion model in Section \ref{sec:contribution}. In Section \ref{sec:exp} we introduce the experimental design and unsupervised baselines. We perform experiments on three IR tasks of the CLEF\footnote{Cross Language Evaluation Forum (http://www.clef-campaign.org/)} evaluation forum and compare our method with typical linear combination methods and more recent unsupervised approaches. The results are presented and discussed in Section \ref{sec:results}.  
Finally, we conclude the paper outlining our findings and contribution in Section \ref{sec:conc}.

\section{Related work}
\label{sec:rel}

In the context of IR, the data fusion task has been casted as the problem of finding a way of merging scores, or ranks, from different output sets, under the assumption that relevant and non-relevant documents show a distinct behavior. Under this assumption, many supervised and unsupervised approaches have been proposed to cope with the data fusion task. 

Probabilistic models have been prominent in the literature. 
A core idea in supervised probabilistic fusion is to assign a relevance probability score to every position in which a document can be returned in the output list, based on the characteristics of the input training set \cite{Lillis2010Estimating}. 

On the other hand, unsupervised score fusion methods under a probabilistic framework have historically relied on Expectation Maximization (EM) algorithms, in order to fit mixture models whose posterior probabilities are aggregated in the fusion stage \cite{Arampatzis2011}. 
However, the fusion performance is limited because the fitting process is static, and the distribution of relevant and non-relevant documents might show a dynamic behavior on a per-query basis. This dynamic behavior has been recently addressed by \cite{LOSADA2018}. They proposed to estimate pseudo-relevance judgments from the ranks of the retrieved documents, and used them to build better score fits for both relevant and non-relevant documents using a log-normal model. Then, a mixture model of two log-normal distributions calculates the posterior probability of each document. Finally, they averaged these probabilities across all IR systems.

Approaches not explicitly modeling score distributions have been historically addressed by linear combination methods of the kind $Comb\langle \textrm{x}\rangle$, with some variants. In the absence of scores, modeling rank distributions has proved to be useful. Rank fusion methods consider the inverse of the rank of each document in each of the output lists as the score. 
Thus, methods such as Reciprocal Rank \cite{zhang2003expansion} or Reciprocal Rank Fusion \cite{cormackRRF2009} boost the signal of top ranked documents, while diminishing it for lower ranks in a nonlinear fashion. 
In the same vein, in  \cite{mouraoISR2014} Mourão and co-authors proposed the Inverse Square Rank (ISR) fusion, which drastically accentuates the nonlinearity of the decaying behavior. With a different objective but producing a similar effect, in \cite{baileyRBC2017} Bailey and colleagues proposed the Rank-Biased Centroid (RBC). They suggested that a geometrically decaying weight function may be used instead of a Borda 
weight to each item. 

Another probabilistic approach is the setting of Archimedean copula functions, which have been proposed for IR under supervised learning. Basically, a copula allows to relate marginal distributions to each other by means of a dependence mechanism \cite{nelsen2013}. In the context of score fusion, Eickhoff et al. trained a Clayton copula to merge scores from a non-relevant training set \cite{eickhoff2013,eickhoff2014}. Once trained, the copula was used to normalize the document scores obtained by the retrieval systems. In this way, the authors proposed two combination methods based on $CombSUM$ and $CombMNZ$, where the fusion mechanism learned by the copula was used as a normalization term of these two combination methods.

Another example is the work by Komatsuda et al. \cite{komatsuda2016}, who proposed a mixture copula as the actual fusion mechanism. Instead of training over a single relevance set, they clustered documents and estimated one copula for each cluster. Then, the authors combined all models by means of a mixture copula, which is a weighted linear combination of the trained models. The weight of each individual copula was calculated as the ratio between the number of relevant documents assigned to a cluster and the total number of relevant documents. 

In this paper, we propose a nonlinear scoring and fusion method that aims to account for IR systems' inter-dependencies and query-to-document correlations. The proposal is supported by the theoretical framework of nested Archimedean copulas, which show several advantages for modeling distinct kinds of data correlations \cite{gorecki_tau_2016}, \cite{nelsen2013},\cite{mcneil2008sampling}.

\section{Problem formalization and objective}
\label{sec:pb}
Our fusion principle is inspired from the theory of copulas. In this section we provide the background of Archimedean copulas in order to formalize our problem and make our assumptions explicit. 

\subsection{Archimedean copulas}
A multivariate distribution is constituted of univariate random variables related to each other by a dependence mechanism. Copulas provide a framework to separate the dependence structure from the marginal distributions. A copula is a multivariate distribution function $C:[0,1]^n \to [0,1]$, defined over the unit cube with uniform margins. 
Sklar's Theorem \cite{sklar1959} gives a way to derive a copula in the general case: 

\begin{theorem}[Sklar] Given $n$ random variables $X_1, \cdots, X_n$ with respective cumulative distribution functions---or marginal cdfs---$F_1, \cdots, F_n$, and joint distribution $H$, there exists a copula $C:\;[0,1]^n \to [0,1]$ such that:
\[
 H(x_1,\cdots,x_n)=C(F_1(x_1),\cdots,F_n(x_n)) 
\]
If the margins are continuous, then this copula is unique. Conversely, if $C$ is a copula and $F_1, \cdots, F_n$ are univariate distribution functions, then $H$ is a multivariate distribution function with margins $F_1, \cdots, F_n$.
\end{theorem}

Of particular interest are Archimedean copulas because they have closed-form expressions with a common method of construction (see \cite{nelsen2013} as a reference textbook).
The bi-variate copula can be constructed according to:
\begin{equation}
	\label{eq:defCopula}
	C(u,v)=\phi\left[\phi^{-1}(u)+\phi^{-1}(v)\right],
\end{equation}
where $\phi:[0, \infty] \to [0,1]$ is a continuous decreasing function called the generator of the copula $C$, such that $\phi(0)=1$, $\phi(\infty):=\lim_{t\to\infty}\phi(t)=0$, and which is strictly decreasing on $[0,\inf\{x:\phi(x)=0\}]$. 

This bi-variate model can be extended to higher dimensions using the so-called exchangeable construction \cite{mcneil2008sampling},\cite{hofert2011sampling}:
\begin{equation}
	\label{eq:defCopula_2d}
	C(\mathbf{u})=\phi(\phi^{-1}(u_1)+\cdots+\phi^{-1}(u_d)), \mathbf{u} \in [0,1]^d,
\end{equation}
provided that $\phi$ fulfills some conditions (see \cite{mcneil2009multivariate} for details). 
Table \ref{tbl:cop} shows most popular one-parameter generators 
defining Archimedean copula families.


\begin{table}[!htbp]
\begin{center}
\begin{minipage}{\textwidth}
\caption{Common Archimedean copula generators and their inverses using one parameter}\label{tbl:cop}
\centering
\begin{tabular}{@{\quad}lc@{}c@{\quad}c}
	\toprule
    \multicolumn{1}{c}{Copula family} &  \multicolumn{1}{c}{$\phi$} & \multicolumn{1}{c}{$\phi^{-1}$} & \multicolumn{1}{@{}l}{Parameter} \\
	\midrule
	Clayton & $(1+x)^{-\frac{1}{\theta}}$ & $x^{-\theta}-1$ & $\theta \in (0,\infty)$ \\
	Frank & $-\frac{1}{\theta}\ln\{1-(1-e^{-\theta})e^{-x}\}$ & $ -\ln \left(\frac{e^{-\theta x}-1}{e^{-\theta}-1} \right)$ & $\theta \in (0,\infty)$ \\
	Gumbel & $e^{-x^{\frac{1}{\theta}}}$ & $(-\ln x)^\theta$ & $\theta \in [1,\infty)$ \\
	Joe & $1-(1-e^{-x})^{\frac{1}{\theta}}$ & $-\ln\left[1-(1-x)^\theta\right] $ & $\theta \in [1,\infty)$\\
	\bottomrule
\end{tabular}
\end{minipage}
\end{center}
\end{table}

The condition of exchangeability implies a dependence structure distributively invariant to permutations, i.e. symmetric. To achieve asymmetric dependence structures, Archimedean copulas can be nested inside each other under certain conditions 
 and allow modeling hierarchical dependence structures.

\subsection{Nested Archimedean copulas as function compositions}

Copulas can be constructed by the nesting of generators \cite{mcneil2008sampling}\cite{joe1997multivariate}. For example, 3-dimensional copulas may be constructed using bi-variate copulas:

\begin{equation}
\label{eq:cop3}
	C(u_1,u_2,u_3)=\phi_1\big(\phi_1^{-1}(u_1)+\phi_1^{-1}\circ\phi_2(\phi_2^{-1}(u_2)+\phi_2^{-1}(u_3))\big)
\end{equation}

 Models like (\ref{eq:cop3}) 
 are called nested or hierarchical copulas. Their iterative construction shows the interaction between generators and inverse functions at each level of the hierarchy. For example, the copula (\ref{eq:cop3}) shows that generators $\phi_1$ and $\phi_2$, with respective parameters $\theta_1$, $\theta_2$, interact in the function composition $\phi_1^{-1}\circ\phi_2$.  It is important to stress however, that in order for Eq.(\ref{eq:cop3}) to be a proper distribution function, we must use the generators of Table \ref{tbl:cop} and satisfy the constraint 
\begin{equation}
	\label{eq:constraint}
	\theta_2 \geq \theta_1;
\end{equation} 
see \cite{mcneil2008sampling} for details. 
Alternatively, copula (\ref{eq:cop3}) 
may be written:
\begin{equation}
\label{eq:cop_3d}
	C(\mathbf{u})=C_1(u_1,C_2(u_2,u_3;\phi_2);\phi_1), \mathbf{u} \in [0,1]^3,
\end{equation}
where the corresponding generators have been made explicit in each copula. 

\begin{figure}[!hptb]
\centering
  \includegraphics[width=0.5\textwidth]{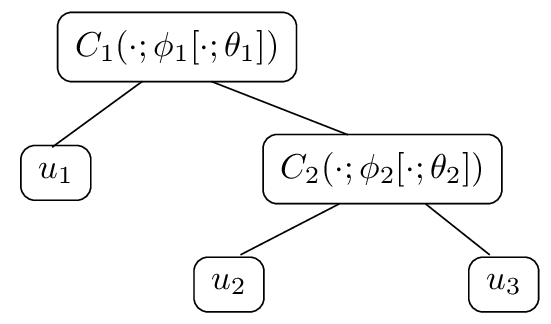} 
\caption{Tree-structure of nested copula (\ref{eq:cop_3d}).}
\label{fig:3d4d-models}
\end{figure}

Nested copulas show a \emph{tree-like} structure allowing for building a hierarchy of nodes, the root node being the copula with generator $\phi_1$. Each node constitutes thus a copula, coupling variables, or copulas, as illustrated in Figure \ref{fig:3d4d-models}, showing the corresponding tree for model  (\ref{eq:cop_3d}).

Different models (with generators from the same family or possibly another family) may be used at each node in the hierarchy \cite{hofert2011sampling}. Indeed, notice that at each level of the hierarchy, there is one generator function operating with its corresponding parameter $\theta$. This is interesting as $\theta$ intuitively captures the correlation between variables when using bi-variate models (e.g. see \cite{mcneil2008sampling}).

The tree structure of nested copulas suggests a data fusion mechanism, from which we have been inspired to propose our nonlinear fusion method. In the context of IR, the intuitive idea is that every observation of a document's rank can be treated as a univariate random variable $U$ so that a hierarchical model 
can be used as a fusion mechanism. We now formalize these intuitions in order to set the problem.

\subsection{Problem formalization and objective}
\label{sec:objective}
Let $\mathcal{D}:=\{d_1,\ldots,d_m\}$ be a document collection; we will assume that we have access to each document's rank $r_j=rank(d_j)$, $j \in \{1,\ldots, m\}$ and $m$ is assumed to be the size of the largest output list. Therefore, for each IR system we can build a result set of rank-document ordered pairs $\mathcal{L}_i := \{ (r_1,d_1)_i , (r_2,d_2)_i , \ldots, (r_m,d_m)_i \}$, $1\leq i\leq n$, where $n$ is the number of IR systems. Alternatively, we may just write $\mathcal{L}_i := \{ r_{i1} , \ldots, r_{im} \}$, where the mapping of document $d_j$ to the rank $r_j$ is implicit. Documents not appearing in some output lists will be appended to those lists. 

The output list $\mathcal{L}_i = \{ r_{i1} , \ldots, r_{im} \}$ of each retrieval system corresponds to a set of measurements from one particular trial, so that each measurement corresponds to a univariate random variable $X$, which is the observation of rank $r_j$ given to document $d_j$ by system $i$. Hence, we assume that each random variable $X$ has a marginal distribution $F_{X}: \mathbb{R} \to [0,1]$, such that 
\[
u_{ij} = F_{X_i}(r_{ij})=P(X_i(d_{ij})\leq r_{ij}),\; u_{ij} \in [0,1]\;\;\;x\in\mathbb{R},
\]
is the probability that document $d_j$ is given at least rank $r_j$ by IR System $i$. In other words, $u_{ij}$ is the \emph{normalized score} of document $d_j$ in output list $i$. 
$F_{X}$ may be estimated using the empirical distribution function $\hat{F}(x)$ of a random variable with $T$ samples $X=\{x_1, x_2, \ldots, x_T\}$:
\begin{equation}
\label{eq:marginals}
\hat{F}(x)=\frac{1}{T+1} \sum_{k=1}^{T} \mathbbm{1}_{(x_k \leq x)}=\frac{\text{number of elements in } X \leq x}{T+1}. 
\end{equation}
In this way, for the $i^{th}$ retrieval system we may compute its result set of normalized scores $U_i:=\{u_{ij}\}$, $j=1,\ldots,m$. 

For each retrieved document, we aim at computing a global score $g_s(\mathbf{u}_j)$ (i.e. system-across), which eventually could translate to a single (fused) rank of $d_j$. 
Let $u_{kj}$ and $u_{(k+1)j}$ be two normalized scores of a same document in respective result sets $\mathcal{L}_k$ and $\mathcal{L}_{(k+1)}$. Our first objective is to compute a nonlinear bi-variate function fusing both scores:
\begin{equation}
\label{eq:fk}
\mathcal{F}_{k}(u_{kj},u_{(k+1)j}; \Theta_k)=\phi_{\theta_{gk}}\big[\phi^*_{\theta_{pk}}(u_{kj})+\phi^*_{\theta_{pk}}(u_{(k+1)j})\big]\;,
\end{equation}
where $\phi$ and $\phi^*$ are non-algebraic functions, and $\Theta_k=(\theta_{gk},\theta_{pk})$ is the parameter set of the bi-variate fusion system. Now, for $n$ IR systems, the overall fusion process is performed by the following function composition that computes a global score for each document $j$:
\begin{equation}
\label{eq:global_score}
g_s(\mathbf{u}_j) = \mathcal{F}_1(u_{1j},\mathcal{F}_2(u_{2j},\mathcal{F}_3(\cdots ,\mathcal{F}_{n-1}(u_{(n-1)j},u_{nj})))),
\end{equation}
where the bi-variate fusion mechanism (\ref{eq:fk}) has been iteratively applied with $k=1,2,\ldots,n-1$. 

\begin{figure}[!hptb]
  \includegraphics[width=0.65\textwidth]{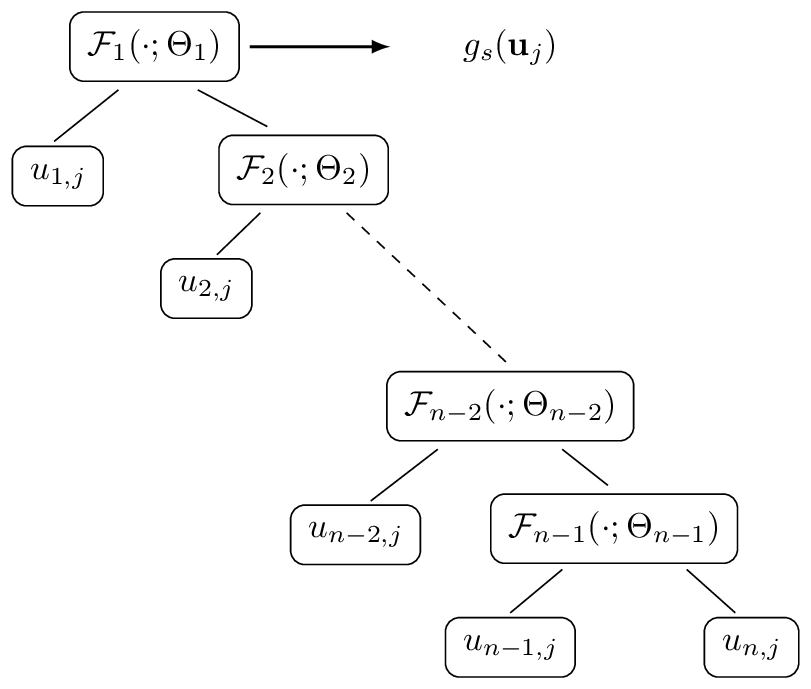} 
\caption{Graphical representation of the fusion process (\ref{eq:global_score}); the model works as a nested $n$-variate copula estimating a global score for a single document $j$ as ranked by $n$ IR systems.}
	\label{fig:fusionSystem}
\end{figure}

At this point, it is worth noticing that expressions (\ref{eq:fk}) and (\ref{eq:global_score}) may be formal copulas. If $\phi^*\equiv \phi^{-1}$ and $\theta_p\equiv \theta_g$, Eq. (\ref{eq:fk}) is a bi-variate Archimedean copula (see Eq. \ref{eq:defCopula}). Moreover, if at each level of the hierarchy defined by Eq. (\ref{eq:global_score}) we have $\theta_{pk}=\theta_{gk}=\theta_k$, and 
if for every pair of functions $\Big(\mathcal{F}_k(\cdot;\theta_k),\mathcal{F}_{k+1}(\cdot;\theta_{k+1})\Big)$ we have $\theta_{k+1} \geq\theta_k$ (see constraint \ref{eq:constraint}), for $k=1,\ldots,n-2$, then Eq. (\ref{eq:global_score}) is a proper nested copula (e.g. like Equation \ref{eq:cop3}). To illustrate this, we will employ the tri-variate case $g_s(\mathbf{u}_j) = \mathcal{F}_1\big(u_{1j},\mathcal{F}_2(u_{2j},u_{3j})\big)$. In this case, the fusion model would be:
\begin{equation}
	\label{eq:fk_3}
	g_s(\mathbf{u}_j) = \phi_{\theta_{g1}}\big[\phi^*_{\theta_{p1}}(u_{1j})+\phi^*_{\theta_{p1}}\circ\phi_{\theta_{g2}}\big(
	\phi^*_{\theta_{p2}}(u_{2j})+\phi^*_{\theta_{p2}}(u_{3j})\big)\big]\;\; \mathbf{u}_j\in[0,1]^3.
\end{equation}
In the framework of copulas, Equation (\ref{eq:fk_3}) may be rewritten:

\begin{equation}
	\label{eq:cop_3}
	g_s(\mathbf{u}_j) = \phi_{\theta_{1}}\big[\phi^{-1}_{\theta_{1}}(u_{1j})+\phi^{-1}_{\theta_{1}}\circ\phi_{\theta_{2}}\big(
	\phi^{-1}_{\theta_{2}}(u_{2j})+\phi^{-1}_{\theta_{2}}(u_{3j})\big)\big]\;\; \mathbf{u}_j\in[0,1]^3,
\end{equation}
which is equivalent to (\ref{eq:cop3}).
If these constraints are not met, then expressions (\ref{eq:fk}) and (\ref{eq:global_score}) are not copulas in a formal sense. Still, both expressions follow the intuition behind the formal definitions of copulas and provide a more general nonlinear framework for data fusion in IR.  
Thus, expression (\ref{eq:global_score}) shows how the score fusion process may be undertaken, by fusing pairs of output lists, using the composition of (\ref{eq:fk}) iteratively. 
Graphically, our proposed fusion process is depicted in Figure \ref{fig:fusionSystem}.

In what follows we describe our proposal in detail. Firstly, we focus the attention on a pragmatic interpretation of the parameters $\theta_g, \theta_p$ and on how we propose to compute them. Then, we describe the fusion technique and the actual implementation of Equations (\ref{eq:fk}) and (\ref{eq:global_score}).

\section{Nonlinear function composition for unsupervised rank fusion in IR}
\label{sec:contribution}
The framework proposed herein is based on fitting nonlinear functions to output data provided by a finite number of retrieval systems.  
Our first hypothesis is that document ranks and scores contain enough evidence to undertake the fusion process over several IR systems. The question is: how to translate the correlations between these relevance indicators to a parameter accounting for their dependence? 
Our second hypothesis is that if we factor each document’s relevance contribution into the parameters of the fusion model, then we may affect the structure of the fusion mechanism to improve performance. In summary, the assumption is that we can take into account a global dependence (or global correlation) between output lists, together with a relevance estimation as per query for each document (or local query-document correlation), for fitting the parameters of the fusion model.
\subsection{A pragmatic interpretation of model parameters}
In order to formalize these intuitions and simplify notation, we propose to rewrite function $\mathcal{F}_k$ (\ref{eq:fk}) as follows:

\begin{equation}
	\label{eq:modelo_bi}
\mathcal{F}(u,v; \theta_g,\theta_p)=\phi_{\theta_g}\big[\phi^*_{\theta_p}(u)+\phi^*_{\theta_p}(v)\big].
\end{equation}

Looking at this equation, the univariate function $\phi^*_{\theta_p}(\cdot)$ can be seen as a nonlinear weighting function (with parameter $\theta_p$) over every document point ($u$ and $v$ for this bi-variate case), and the also univariate function $\phi_{\theta_g}(\cdot)$ can be seen as the combination or fusion process. As discussed in Section \ref{sec:objective}, in the formal setting of copula theory, $\phi^*$ is the inverse generator of $\phi$ (i.e. $\phi^*\equiv\phi^{-1}$) and $\theta_g=\theta_p=\theta$, the sole parameter of the copula (see Equation \ref{eq:defCopula}).
For our purposes, we may relax these constraints at the cost of not being able to obtain a formal copula. In that case, we will keep a correlation between $\theta_g$ and $\theta_p$ as described further below. 

In the setting of copulas, $\theta$ abstracts the correlation between variables, which can be estimated through the Kendall's $\tau$ \cite{nelsen2013}. In our setting, this will be the role of $\theta_g$, and the role of $\theta_p$ will be to \emph{modulate} this correlation by means of an estimation of the relevance of each document across output lists. 

\subsubsection{$\theta_g$ and the correlation between output lists}

Kendall's $\tau$ measures a form of association known as concordance. Two observations $(x_1,y_1)$ and $(x_2,y_2)$ of a pair $(X,Y)$ of random variables are \emph{concordant} if $(x_1-x_2)(y_1-y_2) > 0$---i.e., the ranks of both elements agree---and are \emph{discordant} if $(x_1-x_2)(y_1-y_2) < 0$. 
In light of this, if we take any two IR systems as random variables $X,Y$, then we can give a nice theoretical account on their point-wise probability of dependence by means of the computation of Kendall's $\tau$ over the realizations $(x,y)^T$ of $X$ and $Y$ respectively---i.e., the rank observed for each document in both IR systems. In this case, given $m$ observations, we can compute the sample version of Kendall's $\tau$ noted $\hat{\tau}$ and given by:
	\begin{equation}
		\label{eq:tau_calc}
		\hat{\tau} = \binom{m}{2}^{-1}\hspace{-1em}\sum_{1\leq j_1\leq j_2\leq m}\hspace{-0.25em}\textrm{sgn}((x_{j_1}-x_{j_2})(y_{j_1}-y_{j_2})).
	\end{equation}
	In the setting of copulas, the relation between Kendall's $\tau$ and the parameter $\theta$ depends on the copula family chosen. Table \ref{tbl:tau} shows this relation for some copula families \cite{hofert2012}.

\begin{table}[!htbp]
\begin{center}
\begin{minipage}{\textwidth}
\caption{Typical one-parameter Archimedean copula generators with corresponding Kendall's tau. $D_1(\theta):=\int_{0}^{\theta}t/(\exp(t)-1) dt/ \theta$ is the Debye function of order one.}\label{tbl:tau}
\centering
\begin{tabular}{lc@{\quad}c@{\quad}c}
	\toprule
	\multicolumn{1}{l}{Family} &  \multicolumn{1}{l}{Parameter} & \multicolumn{1}{c}{$\phi$} & \multicolumn{1}{c}{$\tau$} \\
	\midrule
	Clayton & $\theta \in (0,\infty)$ & $(1+x)^{-\frac{1}{\theta}}$ & $\theta/(\theta+2)$ \\
	Gumbel & $\theta \in [1,\infty)$ & $e^{-x^{\frac{1}{\theta}}}$ & $(\theta-1)/\theta$ \\
	Joe & $\theta \in [1,\infty)$ & $1-(1-e^{-x})^{\frac{1}{\theta}}$ & $1-4\Sigma_{k=1}^{\infty}1/(k(\theta k+2)(\theta(k-1))$\\
	Frank & $\theta \in (0,\infty)$  & $-\frac{1}{\theta}\ln\{e^{-x}(e^{-\theta}-1)+1\}$ & $1+4(D_1(\theta)-1)/\theta$\\
	\bottomrule
\end{tabular}
\end{minipage}
\end{center}
\end{table}

For our purposes, Gumbel and Clayton copula families allow for practical closed-form relations. This has a direct implication in the form of our function $\mathcal{F}$ in Eq. (\ref{eq:modelo_bi}) As we mentioned above, $\mathcal{F}$ may be a formal copula. Thus, in order to be able to use copula models as fusion techniques, we will keep Clayton and Gumbel generators as our basic functions. Therefore, we let $\theta_g = \theta$ as shown in Table \ref{tbl:theta}. 

\begin{table}[!htbp]
\begin{center}
\begin{minipage}{\textwidth}
\caption{The form of parameter $\theta_g$ in our fusion framework.}\label{tbl:theta}
\centering
\begin{tabular}{@{}l|l@{}}
	\toprule
	For Clayton based models  $\mathcal{F}$ & $\theta_g = \theta= h(\tau)=2\hat{\tau}/(1-\hat{\tau})$ \\
	For Gumbel based models  $\mathcal{F}$ & $\theta_g  = \theta= h(\tau)=1/(1-\hat{\tau})$ \\ \bottomrule
\end{tabular}
\end{minipage}
\end{center}
\end{table}

\subsubsection{$\theta_p$ and the estimation of relevance}

	The problem turns now to compute $\theta_p$. According to constraint (\ref{eq:constraint}), the condition $\theta_2 \geq \theta_1$ implies a higher concordance between the data of copula 2 with respect to copula 1 (see \cite[p.7]{mcneil2008sampling}). Thus, in a nested copula architecture, the innermost copula takes the pair of random variables with the highest concordance, among all possible pairs. In our setting, this translates to the condition
	\begin{equation}
		\label{eq:constraint_m}
	\theta_{g2} \geq \theta_{p1},
	\end{equation}
	 see Equation (\ref{eq:cop_3}). Recall that if we let $\phi^* \equiv \phi^{-1}$ and set $\theta_{pk} = \theta_{gk} = \theta_k$, our fusion model (\ref{eq:modelo_bi}) becomes a formal copula. In our more general framework, we will let $\theta_p$ to be correlated with $\theta_g$, the cost being that the fusion model (\ref{eq:modelo_bi}) may no longer be a copula. To assign a value to $\theta_p$, we assume that it should modify the correlation structure between pairs of lists according to the relevance of each document.  Thus, to increase the concordance, we propose to increase the value of $\theta_p$ with the documents' relevance $\texttt{rel}_d$ in the following way: 
	\begin{equation}
	\label{eq:point}
	\theta_{pk} = \theta_{gk} \times \texttt{rel}_d,
	\end{equation}
	at each level of the hierarchy, or, in other words, for each pair of output lists.
	To estimate relevance, we will consider the following pieces of information:
	\begin{enumerate}
		\item The query-document match as an indicator of relevance independently of the retrieval
		system used.
		\item The consistency of each document rank across all IR systems.
	\end{enumerate}

	For each document, we take these criteria into account by considering a \emph{query coverage factor} ($\texttt{query}_{\texttt{Cov}}$), defined by a query-document match for each output list normalized by the query length and, a \emph{consistency coefficient} ($\texttt{cons}_{\texttt{IRS}}$), which measures the consistency of each document’s rank among the different output lists:
	
	\begin{align}
	\texttt{query}_{\texttt{Cov}} &= \frac{\texttt{match}(q,d)}{\texttt{length}(q)} \label{eq:doc:match}\\
	\texttt{cons}_{\texttt{IRS}} &= \frac{u_{1d}\times u_{2d}}{u_{1d}+u_{2d}} \label{eq:doc:sim}\\
	\texttt{rel}_d &= \texttt{query}_{\texttt{Cov}} + \texttt{cons}_{\texttt{IRS}}  \label{eq:doc:weight}
	\end{align}
	
	The coefficient $\texttt{cons}_{\texttt{IRS}}$ lowers down the relevance of a document either when it is found at low ranks or when there is a discrepancy of its rank across the IR systems. By making $\texttt{query}_{\texttt{Cov}}$ and $\texttt{cons}_{\texttt{IRS}}$ to range both between $[0,1]$, $\texttt{rel}_d$ would range between $[0,2]$ as defined by Eq. \ref{eq:doc:weight} above. In this way, $\theta_{p}$ could double $\theta_{g}$ if the document is quite relevant or be zero if it is estimated non-relevant. 	

\subsection{Bi-variate copulas as fusion mechanisms}
The first unsupervised fusion models we propose are bi-variate Clayton and Gumbel copulas.  

\vspace{1em}

\hfill\begin{minipage}{\dimexpr\textwidth-1cm}
\begin{description}
	\item[Clayton copula:]	
	\begin{eqnarray*}Generators:\left\{
		\begin{split}
			\phi_{\theta}(u) &= (1+u)^{-1/\theta}\\
			\phi_{\theta}^{-1}(u) &= u^{-\theta}-1
		\end{split}\right.
	\end{eqnarray*}
	
	\begin{equation}
		\label{eq:clayton}
		\mathcal{C}(u,v;\theta) = (u^{-\theta} + v^{-\theta}-1)^{-1/\theta}.
	\end{equation}
\end{description}
	\xdef\tpd{\the\prevdepth}
\end{minipage}
\vspace{1em}
\hfill\begin{minipage}{\dimexpr\textwidth-1cm}
	\begin{description}
		\item[Gumbel copula:]
		\begin{eqnarray*}Generators\left\{
			\begin{split}
				\phi_{\theta}(u) &=\exp\big(-u^{1/\theta}\big) \\
				\phi_{\theta}^{-1}(u) &=\big(-\log(u)\big)^{\theta} \nonumber
			\end{split}\right.
		\end{eqnarray*}
		\begin{equation}
			\label{eq:gumbel}
			\mathcal{G}(u,v;\theta) = \exp\big[-[(-\log(u))^{\theta}+(-\log(v))^{\theta}]^{1/\theta}\big]
		\end{equation}
	\end{description}
	\xdef\tpd{\the\prevdepth}
\end{minipage}

\subsection{Composition of bi-variate non-algebraic functions as fusion mechanisms}
We now propose specific closed-forms for fusion models (\ref{eq:modelo_bi}) inspired by the generators of Clayton and Gumbel copulas respectively. We call them Nonlinear Function Composition (NFC) fusion models.

\vspace{1em}
\hfill\begin{minipage}{\dimexpr\textwidth-1cm}
	\begin{description}
		\item[$\mathbf{F_{PF}}$]: \textbf{NFC based on Power function pairs.} 
		\begin{eqnarray*}Base\,function\,pair\left\{
		\begin{split}
		\phi_{\theta_{g}}(u) &= (1+u)^{-1/\theta_g}\\
		\phi_{\theta_{p}}^{*}(u) &= u^{-\theta_p}-1
		\end{split}\right.
		\end{eqnarray*}
		\begin{eqnarray*}
		\begin{split}
		F_{PF}(u,v) &= \phi_{\theta_{g}}(\phi_{\theta_{p}}^{*}(u)+\phi_{\theta_{p}}^{*}(v)) \\
		&= \phi_{\theta_{g}}(u^{-\theta_p}-1 + v^{-\theta_p}-1) \\
		&= (u^{-\theta_p} + v^{-\theta_p}-1)^{-1/\theta_g}
		\end{split}
		\end{eqnarray*}
	\end{description}
	\xdef\tpd{\the\prevdepth}
\end{minipage}
\vspace{1em}

\noindent Therefore, our first NFC fusion model for two retrieval systems is defined as:
\begin{equation}
	\label{eq:pow}
	F_{PF}(u,v;\theta_g,\theta_p) = (u^{-\theta_p} + v^{-\theta_p}-1)^{-1/\theta_g}
\end{equation}

\hfill\begin{minipage}{\dimexpr\textwidth-1cm}
	\begin{description}
		\item[$\mathbf{F_{EL}}$]: \textbf{NFC based on Exponential-Logarithmic function pairs.} 
		\begin{eqnarray*}Base\,function\,pair\left\{
		\begin{split}
		\phi_{\theta_{g}}(u) &=\exp\big(-u^{1/\theta_{g}}\big) \\
		\phi_{\theta_{p}}^{*}(u) &=\big(-\log(u)\big)^{\theta_{p}} \nonumber
		\end{split}\right.
		\end{eqnarray*}
		\begin{eqnarray*}
		\begin{split}
		F_{EL}(u,v) &= \phi_{\theta_{g}}(\phi_{\theta_{p}}^{*}(u)+\phi_{\theta_{p}}^{*}(v)) \\
		&=\phi_{\theta_{g}}((-\log(u))^{\theta_{p}}+(-\log(v))^{\theta_{p}}) \\
		&=\exp[-((-\log(u))^{\theta_{p}}+(-\log(v))^{\theta_{p}})^{1/\theta_{g}}]
		\end{split}
		\end{eqnarray*}
	\end{description}
	\xdef\tpd{\the\prevdepth}
\end{minipage}
\vspace{1em}

\noindent Therefore, our second NFC fusion model for two retrieval systems is defined as:
\begin{equation}
	\label{eq:exp-log}
	F_{EL}(u,v;\theta_g,\theta_p) = \exp[-((-\log(u))^{\theta_{p}}+(-\log(v))^{\theta_{p}})^{1/\theta_{g}}]
\end{equation}

\subsection{Fusion algorithm}
\label{sec:algorithm}

\begin{figure}[H]
	\centering
	\includegraphics[scale=0.4]{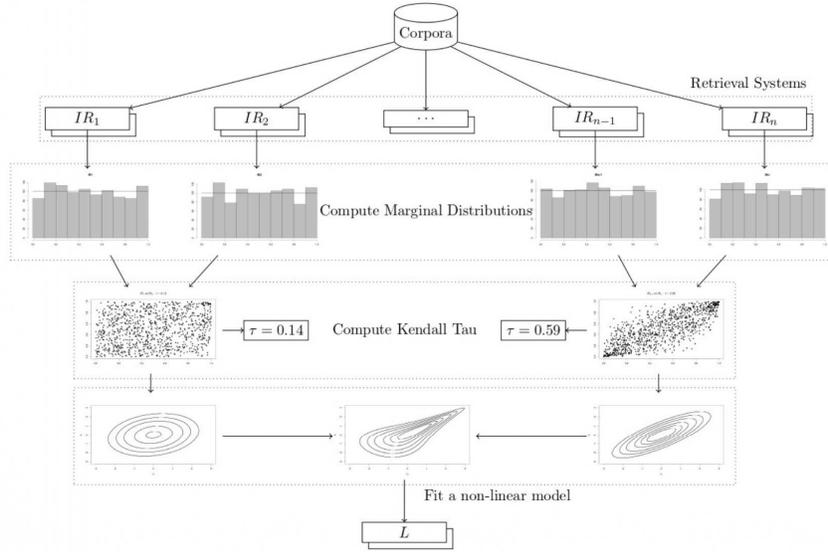}
	\caption{Proposed Unsupervised Nonlinear Function Composition for Rank Fusion framework.}
	\label{fig:fit}
\end{figure}

Our new Unsupervised Nonlinear Function Composition for Rank Fusion framework is depicted in Fig. \ref{fig:fit}.

We call the fusion algorithm 
$Nested\langle\mathcal{F}\rangle$, which provides a framework that exploits a nested architecture of Archimedean copulas, thus performing a fully nonlinear fusion process. In the setting of copulas ---models (\ref{eq:clayton}) and (\ref{eq:gumbel})---, the constraint (\ref{eq:constraint}) must be satisfied for every pair of copulas in the hierarchy; that is, $\theta_{2} \geq \theta_{1}$, where $\theta_{2}$ is the parameter of the innermost copula within the hierarchy. For the non-algebraic function compositions proposed ---models (\ref{eq:pow}) and (\ref{eq:exp-log})---, the constraint (\ref{eq:constraint_m}) must be met for each pair of models; that is, $\theta_{g2} \geq \theta_{p1}$ at every fusion cycle, where $\theta_{g2}$ is the global parameter of the innermost fusion model within the hierarchy.

\begin{algorithm}
    \caption{$Nested\langle\mathcal{F}\rangle$}\label{algNestedC-abstracted}
    \begin{algorithmic}[1]
    \Require {$\mathcal{L}_I$: $n$ IR systems; $\mathcal{F}$: Fusion model; $q$: query string}
    \State $k \gets 0$
    \State $\mathcal{U}_{O}\gets\{1:U_1,\ldots,n:U_n\}$ \Comment{Hashtable of marginals}
    \State $p\gets \lvert\mathcal{U}_{O}\rvert$

    \Repeat
        \State $\ell \gets {p \choose 2}$  
        \State $\mathcal{U}_{\Theta}\gets [(\theta_1,(U,V)_1),\ldots,(\theta_\ell,(U,V)_\ell)]; U,V\in \mathcal{U}_O; \theta_i=h(\tau_i)$\Comment{Using Table \ref{tbl:theta}}
		\State $\mathcal{U}_{{\Theta}_{\texttt{max}}}\gets {\displaystyle \max_{\theta}}(\mathcal{U}_{\Theta})$
		\State $\theta_{current} \gets \mathcal{U}_{{\Theta}_{\texttt{max}}}.\theta$ \Comment{Constraint \ref{eq:constraint}}
		\State $\theta_g \gets \max(0,\theta_{current})$ 
		\State $U,V \gets \mathcal{U}_{{\Theta}_{\texttt{max}}}.(U,V)$
		\State $\mathcal{U}'\gets [\;]$ \Comment{List of pairwise fused scores}
        \ForEach{$j \in \{1,\ldots,m\}$}
			\State $(u,v)\gets(U[j],V[j])$
			\State $d_j \gets U.d_j$
			\If{$\mathcal{F}$ is $Clayton$ or $Gumbel$ copula}
				\State $\theta_{p}\gets\theta_g$
			\Else
				\State $\texttt{cons}_{\texttt{IRS}}\gets\frac{u*v}{u+v}$
				\State $\texttt{query}_{\texttt{\texttt{Cov}}}\gets\frac{\texttt{match}(q,d_j)}{\texttt{length}(q)}$
				\State $\texttt{doc}_{\texttt{weight}}\gets\texttt{query}_{\texttt{Cov}}+\texttt{cons}_{\texttt{IRS}}$
				\State $\theta_{p}\gets\min(\theta_g,\theta_{g}*\texttt{doc}_{\texttt{weight}})$ \Comment{Constraint \ref{eq:constraint_m}}
			\EndIf
			\State $u'_j\gets\mathcal{F}((u,v);\theta_g,\theta_p)$
			\State add $(d_j,u'_j)$ to $\mathcal{U}'$
        \EndFor
		\State Remove $U$ and $V$ from $\mathcal{U}_O$
		\State $p\gets \lvert\mathcal{U}_{O}\rvert+1$
		\State update $\mathcal{U}_O$ by inserting $(p,\mathcal{U}')$
	\Until{$p=1$}
	\Ensure{$\mathcal{U}_O$} \Comment{Fused output list}
\end{algorithmic}
\end{algorithm}

\section{Experimental setting}
\label{sec:exp}
In this section, we describe the experimental setting which includes performance measures, data set, baselines and the concrete set of experiments we carried to assess our fusion models.  
\subsection{Performance measures}
The most frequently used measure for the evaluation of a ranked list is Mean Average Precision (MAP), which gives an overview of its quality by combining both precision and recall and Mean Reciprocal Rank (MRR) also measures the quality of  documents positioned in the first places. 
In line with previous work using copulas for IR \cite{komatsuda2016}, we also use the Precision at $n$ ($P@n$) measure for $n=5,10,20$, which corresponds to the number of relevant results found on the first 5, 10 and 20 positions respectively.

\subsection{Data}
In this work, we used Los Angeles (LA) Times 1994 and Glasgow Herald (GH) 1995 data. Both collections consist in news documents that were part of the CLEF\footnote{Cross Language Evaluation Forum (http://www.clef-campaign.org/)} evaluation forum and are employed for different tasks. 
\subsection{Baselines}
\label{sec:exp:base}
We perform experiments using 3 and 6 IR systems. The IR models we used are (see \cite{IRManningBook} as a reference textbook) 
BM25 
, Vector Space Model (VSM) 
, Language Model (LM) 
, Information Based (IB) Model 
, DFR or divergence from randomness framework 
, and a Language Model using the  Jelinek-Mercer smoothing method --- LMJelinekMercer (LMJM) \cite{zhai2017study}. These IR models were chosen for their ranked lists differ from each other, and because they are available in the Apache Lucene platform used for the experiments. 

As for the score fusion baselines, we chose the following:

\begin{itemize}
\item $CombMNZ$ \cite{fox1994}, as it is a standard in practice in linear combination methods due to its good performance. 
\item $Inverse\;Square\;Rank\;(ISR)$ \cite{mouraoISR2014}, a rank fusion method which drastically accentuates the nonlinearity of the decaying behavior. 
\item $Rank-Biased\;Centroid\;(RBC)$, another state-of-the-art rank fusion method from \cite{baileyRBC2017}.
\item As a final baseline, we will merge the ranks of each possible pair of output lists, both using copulas (Clayton and Gumbel) and using our nonlinear functions ($F_{PF}$ and $F_{EL}$), giving one fused list for every two IR systems in each case. Thence, we proceed to average the pair-wise merged lists in order to obtain a single final list. Finally, we keep the best of those 4 output lists as baseline. This is a typical approach in the literature, whereby a linear combination method is employed to combine scores computed by a nonlinear technique. We call this baseline either $Best_{av}\;C$, $Best_{av}\;G$, $Best_{av}\;F_{PF}$ or $Best_{av}\;F_{EL}$.

\end{itemize}

\subsection{Experiments}
In order to assess the performance of our models (\ref{eq:clayton}) to (\ref{eq:exp-log}), we designed 2 experiments as follows:

\begin{description}
	\item[\textsf{\textbf{Experiment 1: Fusion by nested copulas}}.] For this experiment, we used Algorithm \ref{algNestedC-abstracted} --- $Nested\langle\mathcal{F}\rangle$ --- to assess copulas (\ref{eq:clayton}) and (\ref{eq:gumbel}) in a fully nested architecture configuration, where we meet the constraint (\ref{eq:constraint}). This experiment allows to compare the effect of nesting. 
	\item[\textsf{\textbf{Experiment 2: Fusion by nonlinear function compositions}}.] For this experiment, we used again Algorithm \ref{algNestedC-abstracted}, $Nested\langle\mathcal{F}\rangle$, testing NFC models (\ref{eq:pow}) and (\ref{eq:exp-log}), where we meet constraint (\ref{eq:constraint_m}).
\end{description}

\section{Results and discussion}
\label{sec:results}
In this section, we present and discuss the results of the 2 experiments. Each group of tasks is summarized as follows: $Geo$, $Adhoc$ and $Robust$ refer to Geo CLEF 2008, AdHoc CLEF 2005 and Robust CLEF 2008 topic sets respectively. 

In all results reported, $NC$ and $NG$ refer to Nested Clayton and Nested Gumbel fusion architectures---using models (\ref{eq:clayton}) and (\ref{eq:gumbel}) respectively---, while $NF_{PF}$ and $NF_{EL}$ refer to the nested NFC fusion architectures---using models (\ref{eq:pow}) and (\ref{eq:exp-log}) respectively.

\subsection{Results of Experiment 1: Score fusion by nested copulas}
\label{sec:results:exp2}
Tables \ref{tbl:resEX2-3} and \ref{tbl:resEX2-6} show the results for Experiment 1 using 3 and 6 IR systems respectively. 

\begin{table}[!htbp]
\begin{center}
\begin{minipage}{0.9\textwidth}
\caption{Performance assessment using 3 IR systems in Experiment 1: BM25, VSM and LM.}\label{tbl:resEX2-3}
			\begin{tabular}{c|c|ccc|c|cc}
				\toprule
				Task & Measures & \multicolumn{4}{c}{Baselines} & \multicolumn{2}{c}{Nested copulas}\\
				& & $CombMNZ$ 
				& $ISR$ & $RBC$ & $Best_{av} \text{ } F_{EL}$ & $NC$ & $NG$ \\
				\midrule
				\multirow{2}{*}{Adhoc} & MAP & 0.272 
				& 0.253 & 0.270 & \textbf{0.275} & 0.270 & 0.270\\
				& MRR & 0.529 
				& 0.511 & 0.532 & 0.541 & 0.533 & \textbf{0.549}\\
				& P@5 & 0.376 
				& 0.324 & 0.376 & \textbf{0.412} & 0.376 & 0.392\\
				& P@10 & 0.358 
				& 0.332 & 0.362 & \textbf{0.368} & 0.364 & 0.362\\
				& P@20 & 0.310 
				& 0.298 & 0.312 & \textbf{0.322} & 0.305 & 0.312\\
				\midrule
				\multirow{2}{*}{Robust} & MAP & 0.314 
				& 0.305 & 0.314 & \textbf{0.320} & 0.316 & 0.316\\
				& MRR & 0.520 
				& 0.512 & 0.517 & \textbf{0.534} & 0.519 & 0.524\\
				& P@5 & 0.366 
				& 0.340 & 0.367 & \textbf{0.388} & 0.367 & 0.373\\
				& P@10 & 0.320 
				& 0.310 & \textbf{0.324} & 0.322 & 0.322 & \textbf{0.324}\\
				& P@20 & 0.267 
				& 0.260 & 0.270 & \textbf{0.272} & 0.263 & 0.265\\
				\midrule
				\multirow{2}{*}{Geo} & MAP & 0.141 
				& 0.152 & 0.142 & \textbf{0.155} & 0.138 & 0.138\\
				& MRR & 0.403 
				& 0.397 & 0.424 & \textbf{0.417} & 0.403 & 0.403\\
				& P@5 & 0.225 
				& 0.225 & 0.233 & \textbf{0.250} & 0.225 & 0.225\\
				& P@10 & \textbf{0.183} 
				& 0.171 & \textbf{0.183} & \textbf{0.183} & 0.175 & 0.171\\
				& P@20 & 0.131 
				& 0.125 & 0.135 & 0.131 & \textbf{0.138} & \textbf{0.138}\\
				\bottomrule[1pt]
			\end{tabular}
\footnotetext{The highest observed performance per metric is highlighted using bold typeface. Statistically significant improvements over the linear fusion baseline $CombMNZ$---measured by means of a Paired Sample T-Test---are denoted by a $\dagger$ at $\alpha=0.1$, and by a $\ddagger$ at $\alpha=0.05$.}

\end{minipage}
\end{center}
\end{table}

\begin{table}[!htbp]
\begin{center}
\begin{minipage}{0.9\textwidth}
\caption{Performance assessment using 6 IR systems in Experiment 1.}\label{tbl:resEX2-6}
			\begin{tabular}{c|c|ccc|c|cc}
				\toprule
				Task & Measures & \multicolumn{4}{c}{Baselines} & \multicolumn{2}{c}{Nested copulas}\\
				& & $CombMNZ$ 
				& $ISR$ & $RBC$ & $Best_{av} \text{ } F_{EL}$ & $NC$ & $NG$\\
				\midrule
				\multirow{2}{*}{Adhoc} & MAP & 0.280 
				& 0.266 & 0.277 & 0.281 & 0.287$^{\dagger}$ & \textbf{0.287}$^{\dagger}$\\
				& MRR & 0.550 
				& 0.545 & 0.549 & 0.548 & 0.584 & \textbf{0.590}\\
				& P@5 & 0.408 
				& 0.368 & 0.412 & 0.416 & 0.436$^{\dagger}$ & \textbf{0.448}$^{\dagger\ddagger}$\\
				& P@10 & 0.366 
				& 0.340 & 0.374 & 0.374 & 0.384 & \textbf{0.394}$^{\dagger\ddagger}$\\
				& P@20 & 0.323 
				& 0.306 & 0.322 & 0.331 & 0.329 & \textbf{0.332}\\
				\midrule
				\multirow{2}{*}{Robust} & MAP & \textbf{0.324} 
				& 0.311 & 0.323 & 0.323 & 0.321 & 0.322\\
				& MRR & 0.538 
				& 0.525 & 0.540 & \textbf{0.543} & 0.537 & 0.541\\
				& P@5 & 0.384 
				& 0.361 & 0.382 & 0.390 & 0.395 & \textbf{0.400}$^{\dagger\ddagger}$\\
				& P@10 & 0.329 
				& 0.312 & 0.332 & 0.325 & 0.329 & \textbf{0.333}\\
				& P@20 & 0.273 
				& 0.262 & 0.273 & 0.274 & 0.277 & \textbf{0.278}$^{\dagger}$\\
				\midrule
				\multirow{2}{*}{Geo} & MAP & \textbf{0.164} 
				& 0.155 & 0.162 & 0.161 & 0.156 & 0.157\\
				& MRR & \textbf{0.514} 
				& 0.383 & \textbf{0.514} & 0.465 & 0.424
				& 0.444\\
				& P@5 & \textbf{0.242} 
				& \textbf{0.242} & \textbf{0.242} & \textbf{0.242} & 0.217
				& 0.225\\
				& P@10 & 0.192 
				& 0.183 & 0.188 & \textbf{0.196} & 0.183 & 0.179\\
				& P@20 & 0.133 
				& \textbf{0.140} & 0.129 & 0.129 & 0.133 & 0.133\\
				\bottomrule[1pt]
			\end{tabular}

\footnotetext{The highest observed performance per metric is highlighted using bold typeface. Statistically significant improvements over the linear fusion baseline $CombMNZ$---measured by means of a Paired Sample T-Test---are denoted by a $\dagger$ at $\alpha=0.1$, and by a $\ddagger$ at $\alpha=0.05$.}

\end{minipage}
\end{center}
\end{table}

Table \ref{tbl:resEX2-3}, shows that copula nesting is a fusion scheme that fails to outperform the best pairwise score averaging technique ($Best_{av}\;F_{EL}$). However, it appears to be a slightly more robust method, since for 6 systems (Table \ref{tbl:resEX2-6}) it does outperform the averaging-based model for most of the metrics. Still, the Geo task remains a challenge for nonlinear models as $CombMNZ$ is the winner in the majority of the metrics. 

\subsection{Results of Experiment 2: Score fusion by non-algebraic function compositions}
\label{sec:results:ex3}
Tables \ref{tbl:resEX3-3} and \ref{tbl:resEX3-6} show the results for Experiment 2 using 3 and 6 IR systems respectively. 

\begin{table}[!htbp]
\begin{center}
\begin{minipage}{0.9\textwidth}
\caption{Performance assessment using 3 IR systems in Experiment 2: BM25, VSM and LM.}\label{tbl:resEX3-3}
			\begin{tabular}{c|c|ccc|c|cc}
				\toprule
				Task & Measures & \multicolumn{4}{c}{Baselines} & \multicolumn{2}{c}{Nested NFC}\\
				& & $CombMNZ$ 
				& $ISR$ & $RBC$ & $Best_{av} \text{ } F_{EL}$ & $NF_{PF}$ & $NF_{EL}$\\
				\midrule
				\multirow{2}{*}{Adhoc} & MAP & 0.272 
				& 0.253 & 0.270 & \textbf{0.275}$^{\dagger\ddagger}$ & 0.264 & \textbf{0.272} \\
				& MRR & 0.529 
				& 0.511 & 0.532 & 0.541 & 0.534 & \textbf{0.553} \\
				& P@5 & 0.376 
				& 0.324 & 0.376 & 0.412 & 0.372 & \textbf{0.416}$^{\dagger\ddagger}$ \\
				& P@10 & 0.358 
				& 0.332 & 0.362 & 0.368 & 0.358 & \textbf{0.372}  \\
				& P@20 & 0.310 
				& 0.298 & 0.312 & 0.322 & 0.305 & \textbf{0.325}$^{\dagger\ddagger}$  \\
				\midrule
				\multirow{2}{*}{Robust} & MAP & 0.314 
				& 0.305 & 0.314 & \textbf{0.320} & 0.309 & \textbf{0.320} \\
				& MRR & 0.520 
				& 0.512 & 0.517 & 0.534 & 0.522 & \textbf{0.539} \\
				& P@5 & 0.366 
				& 0.340 & 0.367 & \textbf{0.388} & 0.359 & 0.386$^{\dagger}$ \\
				& P@10 & 0.320 
				& 0.310 & \textbf{0.324} & 0.322 & 0.318 & \textbf{0.324} \\
				& P@20 & 0.267 
				& 0.260 & 0.270 & \textbf{0.272} & 0.262 & 0.271 \\
				\midrule
				\multirow{2}{*}{Geo} & MAP & 0.141 
				& 0.152 & 0.142 & \textbf{0.155} & 0.132
				& 0.143 \\
				& MRR & 0.403 
				& 0.397 & \textbf{0.424} & 0.417 & 0.399 & 0.372 \\
				& P@5 & 0.225 
				& 0.225 & 0.233 & \textbf{0.250} & 0.225 & 0.233 \\
				& P@10 & 0.183 
				& 0.171 & 0.183 & 0.183 & 0.175 & \textbf{0.188} \\
				& P@20 & 0.131 
				& 0.125 & \textbf{0.135} & 0.131 & 0.133 & 0.125 \\
				\bottomrule[1pt]
			\end{tabular}

\footnotetext{The highest observed performance per metric is highlighted using bold typeface. Statistically significant improvements over the linear fusion baseline $CombMNZ$---measured by means of a Paired Sample T-Test---are denoted by a $\dagger$ at $\alpha=0.1$, and by a $\ddagger$ at $\alpha=0.05$.}

\end{minipage}
\end{center}
\end{table}

\begin{table}[!htbp]
\begin{center}
\begin{minipage}{0.9\textwidth}
\caption{Performance assessment using 6 IR systems in Experiment 2.}\label{tbl:resEX3-6}
			\begin{tabular}{c|c|ccc|c|cc}
				\toprule
				Task & Measures & \multicolumn{4}{c}{Baselines} & \multicolumn{2}{c}{Nested NFC}\\
				& & $CombMNZ$ 
				& $ISR$ & $RBC$ & $Best \text{ } NG$ & $NF_{PF}$ & $NF_{EL}$\\
				\midrule
				\multirow{2}{*}{Adhoc} & MAP & 0.280 
				& 0.266 & 0.277 & \textbf{0.287} & 0.279 & 0.284 \\
				& MRR & 0.550 
				& 0.545 & 0.549 & 0.590 & 0.572 & \textbf{0.616}$^{\dagger}$ \\
				& P@5 & 0.408 
				& 0.368 & 0.412 & \textbf{0.448} & 0.424 & 0.444$^{\dagger\ddagger}$ \\
				& P@10 & 0.366 
				& 0.340 & 0.374 & 0.394 & 0.388$^{\dagger\ddagger}$ & \textbf{0.398}$^{\dagger\ddagger}$ \\
				& P@20 & 0.323 
				& 0.306 & 0.322 & 0.332 & \textbf{0.334}$^{\dagger}$ & 0.333 \\
				\midrule
				\multirow{2}{*}{Robust} & MAP & \textbf{0.324} 
				& 0.311 & 0.323 & 0.322 & 0.314
				& 0.323 \\
				& MRR & 0.538 
				& 0.525 & 0.540 & 0.541 & 0.526 & \textbf{0.565} \\
				& P@5 & 0.384 
				& 0.361 & 0.382 & \textbf{0.400} & 0.380 & 0.399 \\
				& P@10 & 0.329 
				& 0.312 & 0.332 & 0.333 & 0.329 & \textbf{0.341}$^{\dagger}$ \\
				& P@20 & 0.273 
				& 0.262 & 0.273 & \textbf{0.278} & 0.275 & 0.275 \\
				\midrule
				\multirow{2}{*}{Geo} & MAP & \textbf{0.164} 
				& 0.155 & 0.162 & 0.157 & 0.158 & 0.152\\
				& MRR & \textbf{0.514} 
				& 0.383 & \textbf{0.514} & 0.444 & 0.431 & 0.389\\
				& P@5 & \textbf{0.242} 
				& \textbf{0.242} & \textbf{0.242} & 0.225 & 0.225 & 0.233 \\
				& P@10 & \textbf{0.192} 
				& 0.183 & 0.188 & 0.179 & 0.179 & 0.171 \\
				& P@20 & 0.133 
				& \textbf{0.140} & 0.129 & 0.133 & 0.131 & 0.119 \\
				\bottomrule[1pt]
			\end{tabular}

\footnotetext{The highest observed performance per metric is highlighted using bold typeface. Statistically significant improvements over the linear fusion baseline $CombMNZ$---measured by means of a Paired Sample T-Test---are denoted by a $\dagger$ at $\alpha=0.1$, and by a $\ddagger$ at $\alpha=0.05$.}

\end{minipage}
\end{center}
\end{table}

The results of Experiment 2 are similar to those of Experiment 1, in that the distribution of winning models for 6 systems is not skewed toward any particular model. We could say, that the nested models perform better in general, except for the Geo task in the 6-system case. Notice that for the 6-system case, we reported the results from model $NG$ of experiment 1 (for the same 6 systems) as the best result baseline $Best\;NG$.

Summarizing these results, we can say that the proposed models, based on the composition of non-algebraic (nonlinear) functions, perform better than the respective copulas that served as inspiration. This is particularly the case of the $NF_{EL}$ technique: algorithm \ref{algNestedC-abstracted} using model (\ref{eq:exp-log}). Still, it must be noted that the pairwise averaging scheme of these composed models yields better results for few systems in general.

One aspect that deserves further discussion is the role of parameter $\theta_p$, as it seems to improve performance almost everywhere with 3 systems, and in the majority of cases for the Adhoc and Robust tasks with 6 IR systems. The Geo task is indeed difficult, and seemed to be better tackled using $CombMNZ$.

\begin{figure}[H]
	\centering
	\begin{subfigure}[b]{\textwidth}
	\centering
		\includegraphics[scale=0.33]{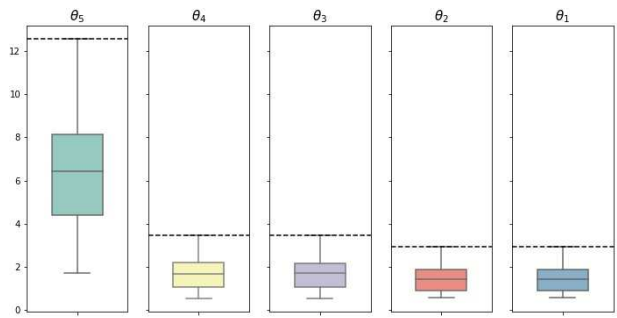}
		\subcaption{$Adhoc$ task.}
		\label{fig:theta_p_a}
	\end{subfigure}
	\begin{subfigure}[b]{\textwidth}
	\centering
		\includegraphics[scale=0.33]{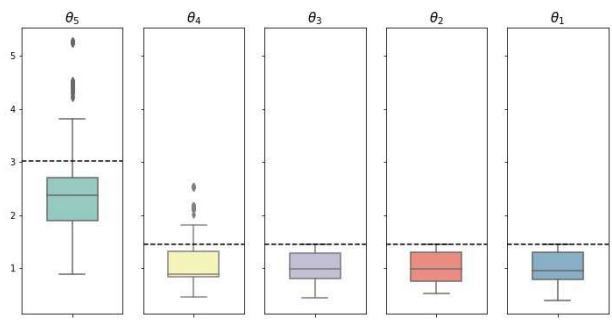}
		\subcaption{$Geo$ task.}
		\label{fig:theta_p_c}
	\end{subfigure}
	\caption{Values of $\theta_g$ (dotted line) and box-plots of $\theta_p$ for the 6-system case (5 fusion cycles), using a fully nested architecture of NFC models. 
	The constraint $\theta_{g_i} \geq \theta_{p_{i-1}}$, can be verified for all $\theta_p$ in each fusion cycle of the Adhoc task and only in some cases for the Geo task.
	}
	\label{fig:plots_theta_p}
	\end{figure}

The parameter $\theta_p$ acts on the score of each document. The constraint (\ref{eq:constraint_m}) forces the score of each item to be affected by a value no greater than $\theta_g$. This can be verified in Figure \ref{fig:plots_theta_p}, where the values of $\theta_g$ and box-plots of $\theta_p$ (around 1000 points ---i.e. retrieved documents) have been plotted for the 6-system case for the Adhoc and Geo tasks. The dotted line corresponds to $\theta_g$ at each fusion step. 

Recall that in the framework of copulas, $\theta_p = \theta_g = \theta$, the sole parameter of the copula. In this case, the constraint (\ref{eq:constraint}) must be met. That is: $\theta_{i} \geq \theta_{i-1}$, must be verified at each fusion cycle; i.e. $\forall i\in\{5,4,3,2\}$ in our 6-system case.

Looking at Figure \ref{fig:plots_theta_p}, it can be readily seen that the corresponding plots of copulas would the same as those depicted in this figure, except that there would be no box-plots. Indeed, there is only one parameter $\theta$ per fusion cycle (corresponding to $\theta_g$ in every case). Looking only at the values of $\theta_g$ in both tasks, we see that the highest values occurs in the $Adhoc$ task.
Thus, according to the interpretation of constraint (\ref{eq:constraint}), we can infer that the better performance, observed in general for the $Adhoc$ task, may be due to those data points being more concordant than the data points in the $Geo$ task.  The low values of $\theta_g$ in $Geo$ task may account for the poorer performance of the copula and NFC models in general for this task. In other words, it seems that the higher the values of Kendall's $\tau$, the better improvement in performance using either nested copulas or our proposed NFC models. Therefore, better fusion systems could be achieved if we use $\tau$ between pairs of IR systems as a decision rule. After some trials using a decision tree, we found that an appropriate threshold of $\tau$ for our proposed NFC models was around $0.4$. In other words, if $\tau \leq 0.4$, it would be better to use a linear fusion mechanism such as $CombMNZ$. 

\begin{figure}[H]
\centering
\includegraphics[width=0.85\linewidth]{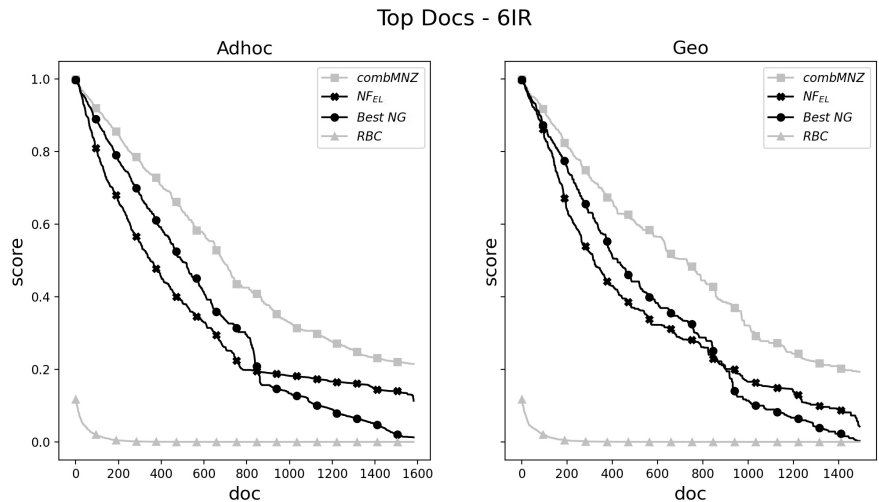}
\caption{Scoring curves for a random query in Experiment 2 using 6 IR systems.}
\label{fig:scoring_EX3}
\end{figure}

Furthermore, in the setting of our NFC models, $\theta_g$ corresponds to the parameter $\theta$ in the framework of copulas, but $\theta_p$ aims at changing the relevance of each document, which translates to the nonlinear way our models score the documents. Given a random query, the nonlinear scoring curve 
for our hierarchical model used in Experiment 2 (Table \ref{tbl:resEX3-6}; model $NF_{EL}$) is shown in Figure  \ref{fig:scoring_EX3} for the Adhoc and Geo tasks. The curves for $CombMNZ$ and $RBC$ baselines are also displayed. Both figures illustrate how our models score non-linearly above $RBC$ but below $CombMNZ$.

\section{Conclusions}
\label{sec:conc}

We proposed an unsupervised data fusion framework based upon non-algebraic function compositions for information retrieval tasks. The method draws from the theory of nested Archimedean copulas, for which supervised methods have been proposed as a way for coping with dependency issues in IR and Data Mining. 

There are three key aspects to our approach. Firstly, we model the output result sets as random variables and use Kendall's $\tau$ as a measure of their concordance. This directly links the correlation between the outcome lists to the dependence parameter of the Clayton and Gumbel Archimedean copulas. The second key aspect is that we cast nested Archimedean copulas as a function composition process. This function composition leads to a hierarchical structure, which allows for modeling asymmetric relationships or data dependencies and is the basis of our fusion strategy. The degree of dependence is captured by the model parameters, which for the chosen copulas may be easily estimated from $\tau$. The third key aspect is that we make explicit the contribution of each document to the change the dependence structure of the model. These three aspects of our approach improve fusion performance when the distributions of the relevant documents do not stray too far from the upper tail. 

Our research constitutes a novel approach on how to use the theory of copulas for score fusion. Future research could include testing other tail dependence copula models, and also change the marginals estimation process in order to fit a well-known distribution that better emulates relevance features. 
For instance, other non-linear normalization techniques include fitting scores into a mix of functions \cite{LOSADA2018}. These techniques could be used as a non-linear marginals estimation process.

\section*{Acknowledgments}
This research was partially supported by the Consejo Nacional de Ciencia y Tecnología (CONACYT), México, through the scholarships numbers:
\begin{itemize}
    \item 296232: F.C. Fernández-Reyes.
    \item 29943: E. Morales-González.
\end{itemize}

\bibliographystyle{unsrt} 
\bibliography{references}

\end{document}